\documentclass[aps,pra,twocolumn,showpacs,amsmath,amssymb,preprintnumbers,superscriptaddress,10pt]{revtex4-1}

\usepackage{amsmath,amssymb}
\usepackage{bm}
\usepackage[latin1]{inputenc}
\usepackage{dcolumn}
\usepackage{graphicx}
\usepackage{SIunits}
\usepackage{hyphenat}

\newcommand{\footnoteremember}[2]{\footnote{#2}\newcounter{#1}\setcounter{#1}{\value{footnote}}}
\newcommand{\footnoterecall}[1]{\footnotemark[\value{#1}]}

\begin{document}

\title{\large{Controlling a superconducting nanowire single-photon detector\\using tailored bright illumination}}

\author{Lars Lydersen}
\email{lars.lydersen@iet.ntnu.no}
\affiliation{Department of Electronics and Telecommunications, Norwegian University of Science and Technology, NO-7491 Trondheim, Norway}
\affiliation{University Graduate Center, NO-2027 Kjeller, Norway}

\author{Mohsen K. Akhlaghi}
\author{A. Hamed Majedi}
\affiliation{Department of Electrical \& Computer Engineering and Institute for Quantum Computing, University of Waterloo, Waterloo, ON, N2L~3G1 Canada}
 
\author{Johannes Skaar}
\author{Vadim Makarov}
\email{makarov@vad1.com}
\affiliation{Department of Electronics and Telecommunications, Norwegian University of Science and Technology, NO-7491 Trondheim, Norway}
\affiliation{University Graduate Center, NO-2027 Kjeller, Norway}

\date{November 16, 2011}

\begin{abstract}
\vspace{1mm}
We experimentally demonstrate that a superconducting nanowire single-photon detector is deterministically controllable by bright illumination. We found that bright light can temporarily make a large fraction of the nanowire length normally-conductive, can extend deadtime after a normal photon detection, and can cause a hotspot formation during the deadtime with a highly nonlinear sensitivity. In result, although based on different physics, the superconducting detector turns out to be controllable by virtually the same techniques as avalanche photodiode detectors. As demonstrated earlier, when such detectors are used in a quantum key distribution system, this allows an eavesdropper to launch a detector control attack to capture the full secret key without being revealed by to many errors in the key.
\vspace{9mm}
\end{abstract}


\maketitle

\section{Introduction}
\label{sec:introduction}

\vspace{1mm}

Quantum key distribution (QKD) allows two parties, Alice and Bob, to generate a secret random key at a distance \cite{bennett1984,ekert1991,gisin2002,scarani2009}. The key is protected by quantum mechanics: an eavesdropper Eve must disturb the signals between Alice and Bob, and therefore reveal her presence. QKD using perfect devices has been proven secure \cite{lo1999,shor2000}.

\vspace{1mm}

Implementations of QKD have to use components available with current technology, which are usually imperfect. While there are numerous security proofs considering more realistic devices \cite{mayers1996,inamori2007,koashi2003,gottesman2004,lo2007,zhao2008a,fung2009,lydersen2010,maroy2010}, these proofs assume that the imperfections are quantified in terms of certain source and detector parameters. Due to the difficulty of characterizing or upper bounding these parameters owing to limitations of these security proofs, it is common to use the more established security proofs for ideal systems also in practical implementations. With actual devices deviating from the ideal models, numerous security loopholes have therefore been identified and usually experimentally confirmed \cite{vakhitov2001,gisin2006,makarov2006,qi2007,zhao2008,lamas-linares2007,fung2007,xu2010,makarov2009,lydersen2010a,jain2011,sun2011}, and in some cases exploited in eavesdropping experiments with full secret key extraction by Eve \cite{gerhardt2011,weier2011}. Finding and eliminating loopholes in implementations is crucial to obtain provable practical security.

\vspace{1mm}

As an example, several recent attacks have been based on bright-light control of avalanche photodiodes (APDs) \cite{makarov2009,lydersen2010a,gerhardt2011,weier2011,lydersen2011,wiechers2011,sauge2011,lydersen2010b,lydersen2011b}. Superconducting nanowire single-photon detectors (SNSPDs) studied in this paper are based on different physics. However, as we will see, the principles of attacks on QKD systems using SNSPDs are broadly similar to attacks on QKD systems using APDs: Eve uses a faked-state attack \cite{makarov2005}, can blind the detectors \cite{makarov2009,lydersen2010a}, make them click with a classical threshold using a bright pulse \cite{lydersen2010a} or let one detector temporarily recover from blinding \cite{makarov2009}; also, detector's response to multiphoton pulses can be superlinear \cite{lydersen2011b}. We refer to these principles through the paper.

\begin{figure*}[t]
   \includegraphics[width=16cm]{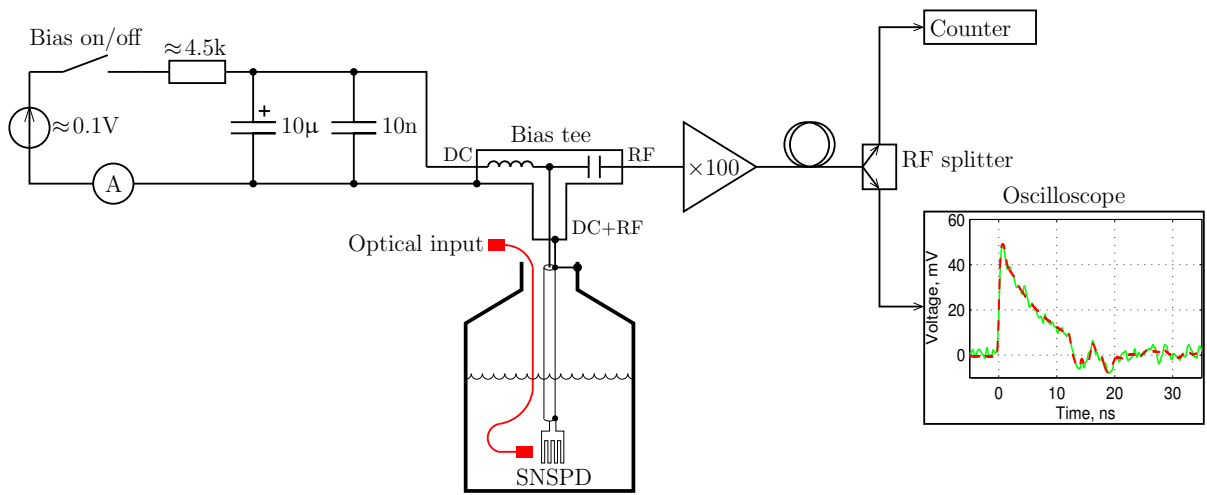}
   \caption{(Color online) Detector circuit. The SNSPD is biased from a battery-powered direct current (DC) source, an equivalent circuit diagram of which is shown. Pulses produced by the SNSPD travel through $\sim 1\,\meter$ coaxial cable, bias tee ($0.1$--$6000\,\mega\hertz$, Mini-Circuits ZFBT-6GW+), radio-frequency (RF) amplifier (voltage gain 100, $0.1$--$1500\,\mega\hertz$, Phillips Scientific 6954-S-100), $\sim 1.5\,\meter$ coaxial cable, and RF splitter (Mini-Circuits ZN2PD-9G-S+), to the counter and oscilloscope. Inside the oscilloscope box: normal single-photon response after the RF amplifier and splitter, shown as a single-shot trace with $2\,\giga\hertz$ bandwidth (green solid line) and averaged over many pulses (red dashed line). Features appearing $12\,\nano\second$ after the leading edge are attributed to reflections due to impedance mismatch in the RF circuits.}
   \label{fig:detector}
\end{figure*}

Although SNSPDs have been used in several QKD experiments \cite{hadfield2006,takesue2007,stucki2009,rosenberg2009,liu2010}, this detector technology is still in its infancy. No automated unattended operation of systems containing SNSPDs has been reported. Technical aspects of SNSPD operation, such as handling the latching behavior and converting the nanowire analog response into a digital detection signal, have only been studied in the normal single-photon counting regime. So far, no attempt has been reported to consider SNSPD's non-idealities in order to attack a QKD system. This study thus serves as an \emph{early warning.} Although we have done our experiments on only one detector sample, we show that control by bright light can be achieved through two separate mechanisms, and may thus be applicable to different detector designs \footnoteremember{fn:testing-restricted-to-one-sample}{Testing was restricted to one sample owing to difficulty of gaining access to more samples. One reason for this was that high-power illumination was initially assumed to be potentially lethal to the devices. However, the sample we tested survived undamaged. In fact, the measurement in Sec.~\ref{sec:control-in-latched-state-physics} suggests that the maximum temperature of the nanowire under $20\,\milli\watt$, $1550\,\nano\meter$ continuous-wave illumination stayed relatively low, because only a part of the nanowire rose above the superconducting transition temperature of $\sim 10\,\kelvin$.}. Regardless of whether the control mechanisms we have identified apply to other detector designs, our experiment shows that the bright illumination response of the SNSPD is deviating from the detector model in the simple security proofs for QKD. Therefore, theoretical and/or experimental effort is required to re-establish security for QKD systems using SNSPDs.

The paper is organized as follows. In Sec.~\ref{sec:detector-description}, we describe the SNSPD under test. Sections~\ref{sec:control-in-latched-state} and \ref{sec:control-via-deadtime-extension} deal with the SNSPD in the latched and non-latched states; in each section we present the physics behind detector's reaction to bright-light illumination, then how it can be exploited to attack QKD. We discuss our findings and conclude in Sec.~\ref{sec:conclusion}.

\section{Detector design and operation}
\label{sec:detector-description}
We performed our tests on an SNSPD of a fairly standard configuration, which has been characterized in previous publications \cite{akhlaghi2008,orgiazzi2009,yan2009}. The SNSPD chip was manufactured by Scontel, Moscow, and consists of a $4\,\nano\meter$ thick, $120\,\nano\meter$ wide NbN nanowire on sapphire substrate, laid out in a $10 \times 10\,\micro\meter$ meander pattern with 60\% filling ratio. The chip is packaged and installed in a $\sim 1\,\meter$ long dipstick assembly (see Ref.~\cite{orgiazzi2009} for details), lowered into a Dewar flask. During detector operation, the chip is immersed into liquid helium at $4.2\,\kelvin$. It is optically accessible through a single-mode fibre. The chip is connected to a room-temperature bias tee and wideband radio-frequency (RF) amplifier via a $50\,\ohm$ coaxial cable (Fig.~\ref{fig:detector}). A battery-powered current source 
biases the superconducting nanowire with $I_\text{b} = 22.5\,\micro\ampere$ which is $\approx 0.85$ of its critical current $I_\text{c}$ (this $I_\text{b}$ value provides the highest ratio of photon detection probability at $1550\,\nano\meter$ to dark count rate, for this particular SNSPD sample). The signal from the output of the RF amplifier is split to a $16\,\giga\hertz$ single-shot oscilloscope (Tektronix DSA 71604) and a counter (Stanford Research Systems SR400). Detection efficiency for single photons at $1550\,\nano\meter$ was $2.2 \times 10^{-5}$ and the dark count rate was $< 1\,\hertz$, which is a typical performance for this SNSPD model (higher detection efficiency can be obtained at the expense of much higher dark count rate; while this SNSPD was not optimised for high detection efficiency, the effects we have observed should qualitatively be the same as with efficiency-optimised designs \cite{rosfjord2006}). The detector sensitivity was polarization-dependent; in all experiments in this paper polarization was aligned to maximize the detection efficiency, using a fiber polarization controller.

One aspect of detector operation is how the analog pulse produced by a transient hotspot (see inset in Fig.~\ref{fig:detector}) is converted into a detection event and assigned a particular timing. The analog pulse is well-defined, its magnitude and shape being nearly constant from one photon detection to another. Therefore almost any discriminator design would work for single-photon detection, and its implementation details (bandwidth, hysteresis, whether it is a threshold discriminator or a constant-fraction discriminator, etc.) are often omitted in the literature on SNSPDs. However, as previously discussed for APDs \cite{yuan2010,lydersen2010c}, these details become more important for demonstration of detector control by bright light. We assume in this study that the analog pulse is sensed by a high-speed voltage comparator, and the detection event timing is registered by pulse's leading edge crossing a pre-set comparator threshold. Indeed this is how our SR400 counter operates: it has an adjustable threshold set with $0.2\,\milli\volt$ resolution. In our setup, the counter works correctly (registering one count per one single-photon analog pulse) in a wide range of threshold settings, $+4.4$ to $+37\,\milli\volt$. A detail not mentioned in the literature is what threshold level the comparator should be set at, within this working range. While the setting may not affect normal detector operation, only a part of this voltage range is reachable under bright-light control described in the following section.

Another interesting aspect of detector operation is latching. In single-photon detection regime, the hotspot after formation shrinks quickly and the nanowire returns to the superconducting state \cite{yang2007}. However the detector also has a stable \emph{latched state,} when a larger self-heating hotspot persists indefinitely, at a steady current $I_\text{latched}$ which is a fraction of $I_\text{b}$, and a large voltage across the SNSPD. The detector is blind to single photons and does not produce dark counts in this regime. A properly designed SNSPD does not enter the latched state after a single-photon detection \cite{yang2007,annunziata2010}. However it can still latch after an electromagnetic interference (which in our experiment was easily caused by switching on and off lights and other mains-powered electrical equipment in the same building). Latching also occurs after a brief bright illumination: as little as $50\,\nano\watt$, $5\,\milli\second$ long single light pulse at $1550\,\nano\meter$ reliably latches the device. Increasing the bias current $I_\text{b}$ very close to $I_\text{c}$ also leads to latching. The only way to return the detector from the latched state into the normal regime is to temporarily reduce $I_\text{b}$ below $I_\text{latched}$. In our experiment, and supposedly in most other experiments reported in the literature, this was performed manually.

\section{Detector control in latched state}
\label{sec:control-in-latched-state}

\subsection{Physics}
\label{sec:control-in-latched-state-physics}
In the latched state, the Joule heat generated in the normally-conductive fraction of the nanowire exactly balances the cooling. The length of the normally-conductive fraction changes with the voltage applied across the SNSPD. We investigated this by replacing the battery-powered bias source with an external bias source consisting of a constant-current source limited at a certain maximum voltage. Since the SNSPD enters and maintains latching at a current lower than the normal bias current, this bias source automatically turns into a voltage source once the device latches. In our experiment, $I_\text{latched}$ was roughly $7\,\micro\ampere$ regardless of the voltage across the device, up to $10\,\volt$ (we did not apply higher voltages to reduce the chance of electrical breakdown). At $10\,\volt$, the nanowire resistance was thus $\sim 1.4\,\mega\ohm$. Above the superconducting transition temperature the resistance of the entire device is approximately constant, and is $\approx 2.3\,\mega\ohm$ \cite{yan2009}. Therefore we concluded that slightly over half its length was normally-conductive at $10\,\volt$. During the experiment, $I_\text{latched}$ would randomly assume a value in the $6$ to $8\,\micro\ampere$ range, which could correspond to the normally-conductive region shifting and ``locking'' to the local variations of nanowire thermal characteristics along its length.

\begin{figure}
   \includegraphics[width=8.6cm]{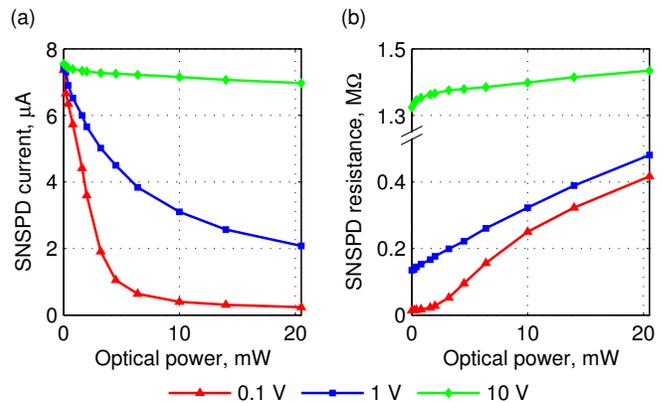}
   \caption{(Color online) Response to continuous-wave (CW) light in the latched state. (a)~Current $I$ through the SNSPD vs.~optical power at $1550\,\nano\meter$, at different voltages $V$ applied across the SNSPD. (b)~SNSPD resistance $R = V/I$.}
   \label{fig:cw-response}
\end{figure}

Next, we investigated what happened when bright continuous-wave (CW) light was applied in the latched state. Under illumination, current $I$ through the device dropped, with a different sensitivity at different voltages (Fig.~\ref{fig:cw-response}(a)). When recalculated into device resistance (Fig.~\ref{fig:cw-response}(b)), we see that at low source voltages the resistance increased by about the same amount ($350$--$400\,\kilo\ohm$ per $20\,\milli\watt$), while at $10\,\volt$ the increase was smaller ($\sim\!\!\!110\,\kilo\ohm$). Note that depending on optical coupling, illumination may be unevenly distributed along the nanowire.

Implementation and maximum voltage of the bias source is yet another detail that varies between setups and is rarely specified in the literature. In our detector it is implemented as a $\approx 0.1\,\volt$ voltage source in series with $\approx 4.5\,\kilo\ohm$ resistor (see Fig.~\ref{fig:detector}), with both voltage and resistance being trimmable in a small range to set precise $I_\text{b}$ in the normal (non-latched) regime. When the SNSPD resistance is zero, this bias circuit acts as a current source. However, in the latched state the SNSPD resistance becomes larger than the circuit output impedance, thus it acts as a voltage source. Measurements done with this battery-powered bias circuit closely match the $0.1\,\volt$ curve in Fig.~\ref{fig:cw-response}.

\subsection{Exploit}
\label{sec:control-in-latched-state-exploit}
The eavesdropper Eve can latch the device by applying sufficient illumination at the SNSPD, for instance a single $> 50\,\nano\watt$, $5\,\milli\second$ long light pulse at $1550\,\nano\meter$. The latching causes a number of random detection events, depending on the discriminator setting and optical power of the latching pulse. However, for intense illumination, it is possible for Eve to latch the device with only a few random events (for instance using $5\,\milli\watt$ optical power for some $\milli\second$ at $20\,\milli\volt$ discriminator threshold). Also note that Eve only have to latch the device once.

\begin{figure}
   \includegraphics[width=8.6cm]{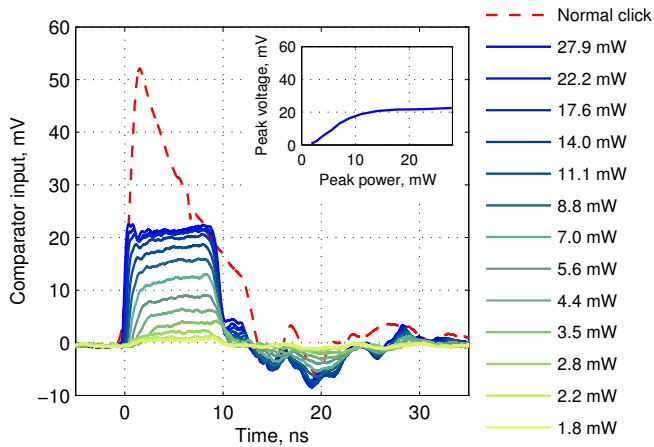}
   \caption{(Color online) Electrical response in the latched state to the $10\,\nano\second$, $1550\,\nano\meter$ optical trigger pulse. All traces are averaged over 500 samples. The trigger pulse saturates at an electrical response of about $20\,\milli\volt$ (see inset), compared to the normal detection event which reaches peak amplitude of about $50\,\milli\volt$.}
   \label{fig:latched-trigger-pulse}
\end{figure}

\begin{figure}
   \includegraphics[width=8.6cm]{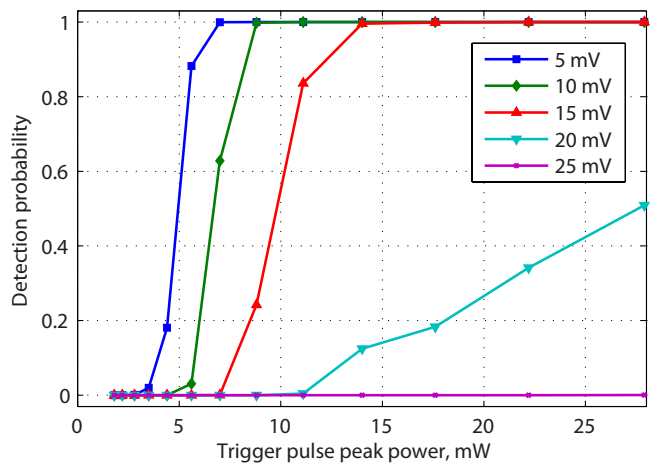}
   \caption{(Color online) Detection probability of the $10\,\nano\second$ trigger pulse, depending on the comparator threshold. The probabilities were obtained by simulating a bandwidth-limited ideal comparator, requiring that the wideband signal recorded by the oscilloscope spent at least $3\,\nano\second$ above the threshold level to register a click. A measurement using a real comparator SR400 (not shown on the plot) confirmed this strong superlinearity; jitter of SR400 comparator click in response to the bright pulse was $\sim 0.5\,\nano\second$ full width at half maximum (FWHM).}
   \label{fig:latched-detection-probability}
\end{figure}

In the latched state, the SNSPD is insensitive to single photons and produces no dark counts (similarly to blinding of APDs \cite{makarov2009,lydersen2010a}). However, the nanowire's response to bright CW illumination detailed in Sec.~\ref{sec:control-in-latched-state-physics} also holds on a nanosecond scale for bright pulses, and can be used to produce an electrical pulse after the RF amplifier and splitter (Fig.~\ref{fig:latched-trigger-pulse}) \footnote{The measurement in Figures~\ref{fig:latched-trigger-pulse} and \ref{fig:latched-detection-probability} was taken some days apart from the measurement in Fig.~\ref{fig:cw-response}. Unfortunately it seems we had an optical and/or polarization alignment problem during the former: the detector is $\sim 3$ times less sensitive in Figures~\ref{fig:latched-trigger-pulse} and \ref{fig:latched-detection-probability} than in the rest of the paper, with its behaviour being otherwise consistent except for the scale factor of $\sim 3$ on the optical power.}. The response is caused by a larger piece of the nanowire becoming normally conductive during the bright illumination, therefore causing an abrupt change in the resistance, just as a single photon causes an abrupt change in the resistance in the normal operating regime. Note that the electrical response to a bright trigger pulse saturates at $\sim 20\,\milli\volt$ when optical power $> 15\,\milli\watt$ is applied, because at this power the current through the nanowire is reduced to almost zero.

Since this analog electrical pulse is sensed by a comparator, the detector has a highly superlinear detection probability of bright pulses \cite{lydersen2011b}. By simulating an ideal bandwidth-limited comparator on recorded wideband long oscilloscope traces, we find that the detection probability would depend strongly on the comparator threshold (Fig.~\ref{fig:latched-detection-probability}). With the comparator threshold in the 5--20$\,\milli\volt$ range, the detection probability is highly superlinear and increases quickly from negligible to a substantial value for a $3\,\deci\bel$ increase in the optical power. A sufficient condition for a detector control attack is a large ratio of detection probabilities over a $3\,\deci\bel$ change in the trigger pulse power \cite{lydersen2010a,lydersen2011b} (or $6\,\deci\bel$ change in the trigger pulse power for distributed-phase-reference protocols \cite{lydersen2011}). Then Eve can intercept the quantum states from Alice, and resend bright trigger pulses corresponding to her detection to Bob \cite{lydersen2010a,lydersen2011b}. If Eve used a measurement basis not matching Bob's, she wants her pulse to remain undetected. Indeed when the pulse is measured by Bob in a different basis, it will be split to both detectors, corresponding to $3\,\deci\bel$ reduction in its power, and almost never cause a click. Due to the large difference in detection probability for $3\,\deci\bel$ change in the trigger pulse amplitude, a detector control attack would cause negligible errors and not expose eavesdropping, for the comparator threshold settings $\lesssim 20\,\milli\volt$. Above $\sim 20\,\milli\volt$ the trigger pulses stop causing clicks at all, and this attack method no longer works. However, it may be possible to reach higher threshold settings using a different attack method described in the next section.

\begin{figure*}
   \includegraphics[width=14cm]{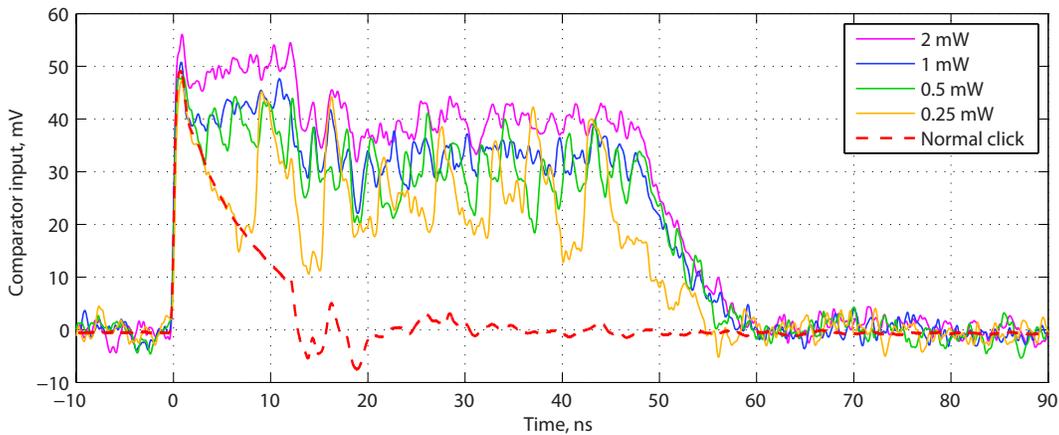}
   \caption{(Color online) Electrical response in the non-latched state to the $48\,\nano\second$, $1550\,\nano\meter$ optical pulse. Single-shot traces with $2\,\giga\hertz$ bandwidth for different pulse powers are shown, as well as an averaged normal single-photon response.}
   \label{fig:deadtime-extension-prepulse}
\end{figure*}

\section{Detector control via deadtime extension}
\label{sec:control-via-deadtime-extension}

\subsection{Physics}
\label{sec:control-via-deadtime-extension-physics}
In this section we consider a non-latched, single-photon sensitive normally operating detector. The attack is based on detector's ability to form a hotspot in response to bright light when the current $I$ through the SNSPD is low. In addition, the hotspot formation probability at a low current is strongly superlinear. It is well-known that at relatively low values of the bias current $I_\text{b}$, multiphoton processes dominate the detector sensitivity \cite{verevkin2002,akhlaghi2009,lydersen2011b}. Here we demonstrate that this effect becomes extreme during the normal recovery time after a photon detection.

\begin{figure}
   \includegraphics[width=8.6cm]{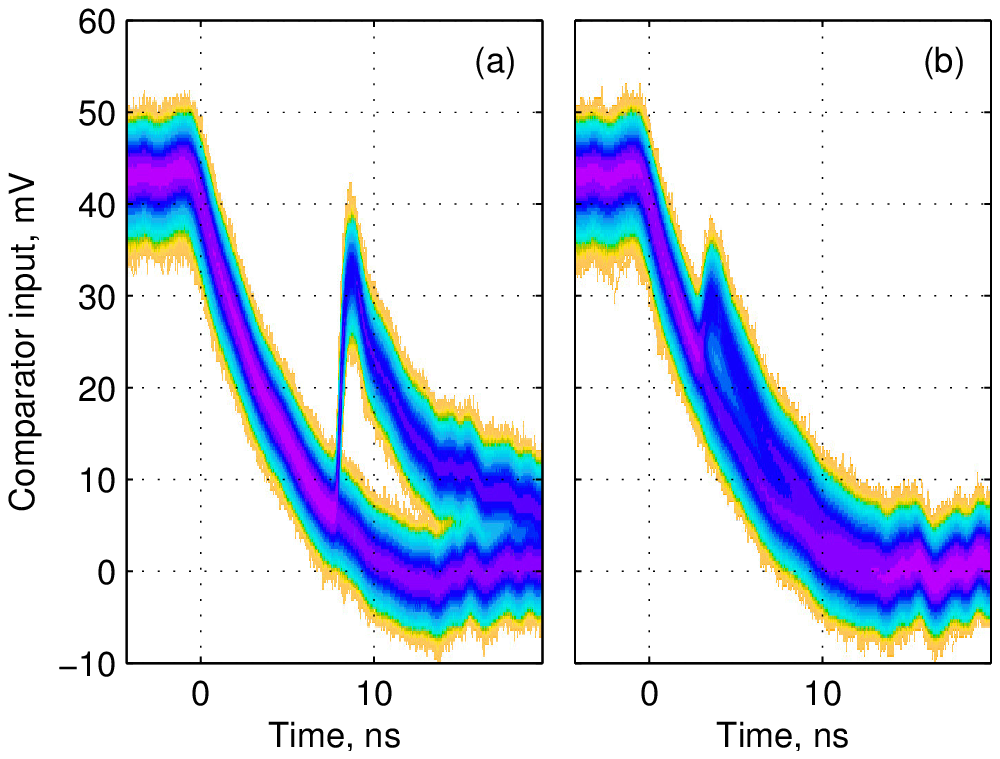}
   \caption{(Color online) Accumulated 30,000 oscilloscope traces of the electrical response to the trigger pulse during the recovery from a $48\,\nano\second$, $2.5\,\milli\watt$ rectangular optical pulse. The trigger is (a) $8\,\nano\second$ into the recovery, 25$\,\femto\joule$ energy, (b) $3\,\nano\second$ into the recovery, 78$\,\femto\joule$ energy. In both cases the trigger pulse causes hotspot formation with roughly $50\%$ probability, and resets the voltage to the same level. All oscillograms at trigger pulse delays $\geq 2\,\nano\second$ show the same behavior.}
   \label{fig:hotspot-during-recovery}
\end{figure}

In normal detector operation, after the hotspot formation, $I$ drops to a fraction of $I_\text{b}$ \cite{yang2007}. Then, $I$ exponentially recovers to $I_\text{b}$ at a slow rate, owing to a relatively large kinetic inductance of the superconducting nanowire (see dashed trace in Fig.~\ref{fig:deadtime-extension-prepulse}). During the initial part of this recovery, the SNSPD remains insensitive to single photons, but it can react to a bright illumination by forming another hotspot, with a higher illumination power being able to form a hotspot earlier in the recovery. This is illustrated in Fig.~\ref{fig:deadtime-extension-prepulse}, which shows electrical response to a $48\,\nano\second$ long bright pulse. At $0.25\,\milli\watt$ pulse power, the single-shot trace clearly shows that the SNSPD forms a hotspot on average every $6\,\nano\second$. At $0.5\,\milli\watt$, the period reduces to $\sim 2.7\,\nano\second$. At higher optical powers separate hotspot formations are no longer distinguishable, but the whole electrical pulse gets higher, indicating a lower average current through the nanowire during the optical pulse. Thus, during a sufficiently bright optical pulse, the electrical signal will stay above the comparator threshold. This allows Eve to extend the detector deadtime after the first photon detection, up to $500\,\nano\second$ with this detector setup, without causing latching.

\begin{figure}
   \includegraphics[width=8.6cm]{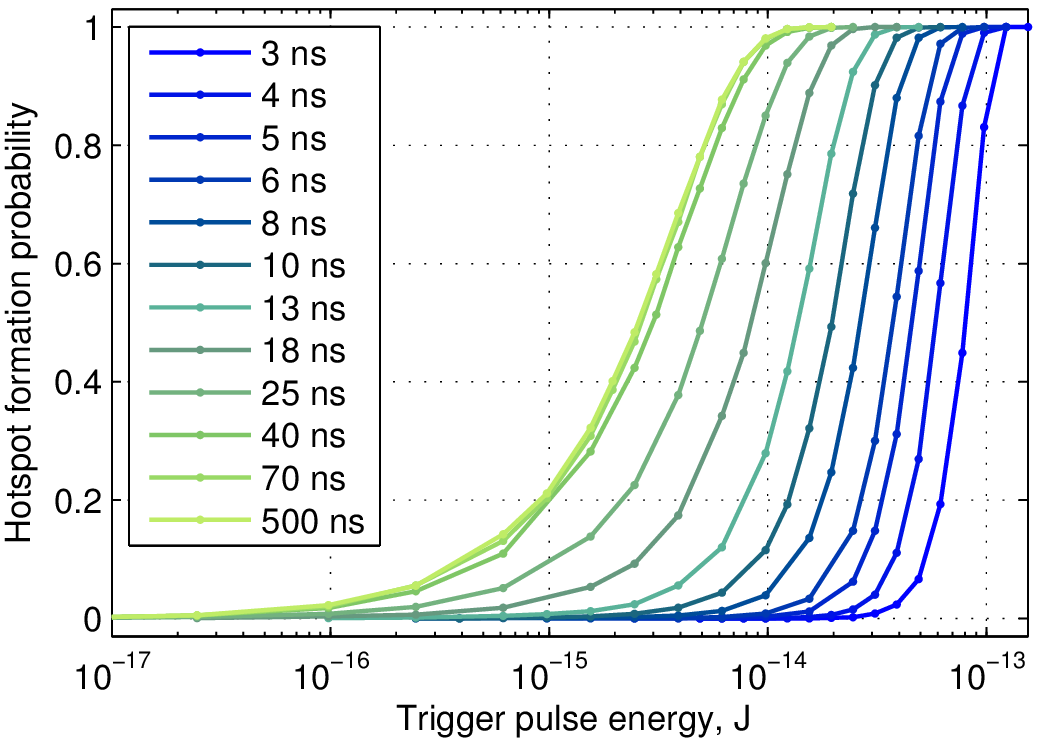}
   \caption{(Color online) Hotspot formation probability vs.\ energy of a $53\,\pico\second$ wide trigger pulse, for different trigger pulse delays after the closing edge of a $48\,\nano\second$, $2.5\,\milli\watt$ rectangular optical pulse (both pulses at $1550\,\nano\meter$). The probabilities were extracted from recorded oscillograms similar to those shown in Fig.~\ref{fig:hotspot-during-recovery}. $10^{-13}\,\joule$ corresponds to 780,000 photons contained in the trigger pulse.}
   \label{fig:deadtime-extension-detection-probabilities}
\end{figure}

We further quantify the hotspot formation probability during the recovery, by applying a $53\,\pico\second$ FWHM trigger pulse after the closing edge of the $48\,\nano\second$, $2.5\,\milli\watt$ pulse. (The recovery after the bright pulse should be similar to the recovery after a single-photon detection, however we focus on the former for reasons that will become apparent in the next subsection.) As far as we can see, response to this trigger pulse is probabilistic and binary: the hotspot either forms, or it does not (Fig.~\ref{fig:hotspot-during-recovery}). In the former case the recovery resets and starts anew from a certain current value, in the latter case the recovery continues undisturbed. The probability that the trigger pulse causes a hotspot is plotted in Fig.~\ref{fig:deadtime-extension-detection-probabilities}. The measurement shows that the detection probability is reduced for at least $40\,\nano\second$. It also shows that the detector is highly superlinear in at least the first $10\,\nano\second$. During this time, a hotspot can be formed with unity probability using a sufficiently high-energy trigger pulse ($\sim 150\,\femto\joule$), while the same trigger pulse attenuated by $20\,\deci\bel$ (i.e., 100 times lower pulse energy) is very unlikely to cause a hotspot formation.

\begin{figure*}[t]
   \includegraphics[width=\textwidth]{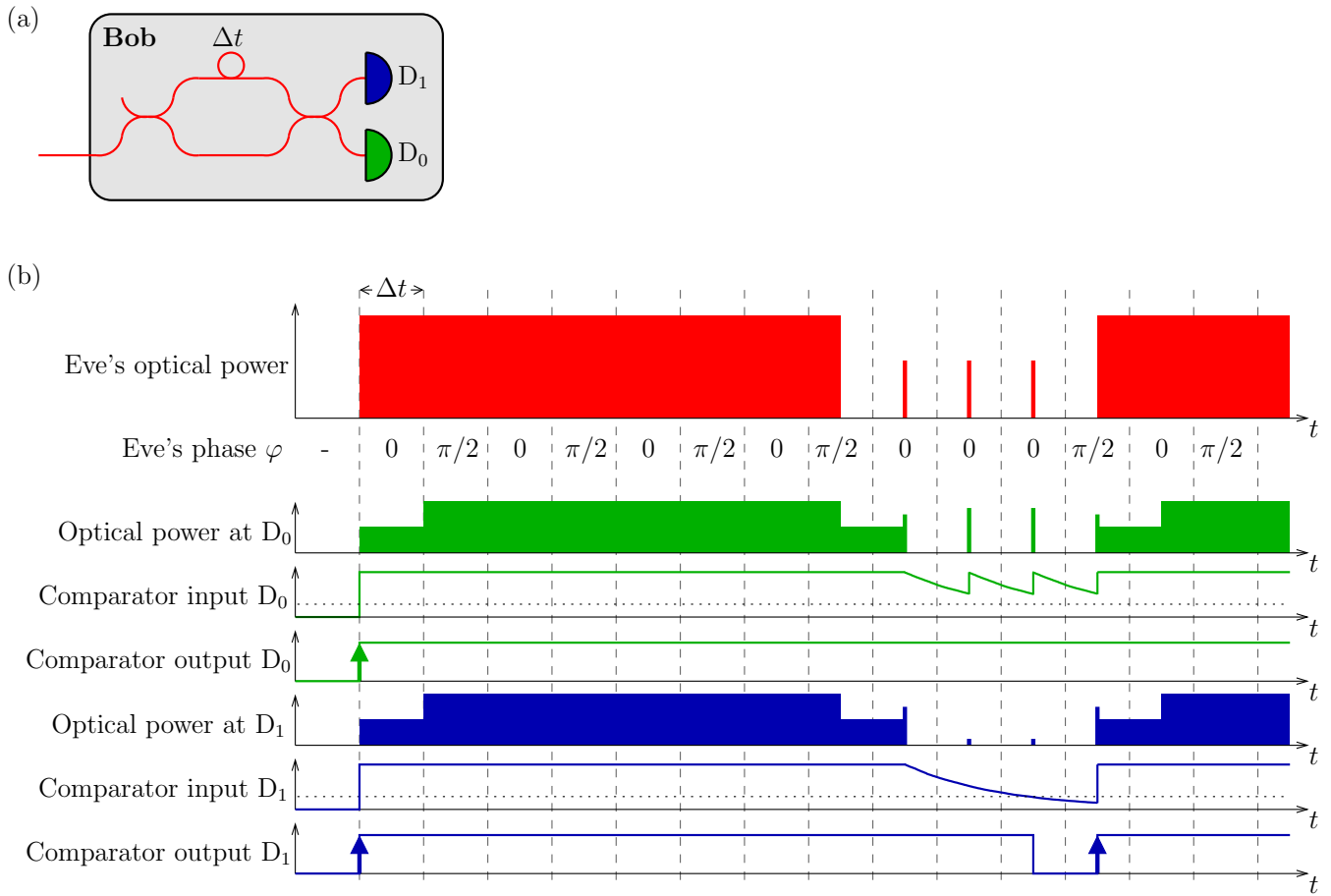}
   \caption{(Color online) Proposed faked-state attack on the DPS-QKD system. (a) Bob's optical scheme. $\Delta t$ is the time delay between the two interferometer arms. (b) Diagram showing Eve's optical output, how her light splits between the two Bob's detectors, and how the electrical signals in each detector react to it.}
   \label{fig:dps-bob-control}
\end{figure*}

\subsection{Exploit}
\label{sec:control-via-deadtime-extension-exploit}
Extendability of SNSPD's deadtime can be exploited in the earlier described attack \cite{makarov2009} on the Bennett-Brassard 1984 (BB84) and similar protocols. We remark that the superlinearity is not required for this attack, but is helpful and makes it easier. Here we propose a version of this attack for differential-phase-shift QKD (DPS-QKD) systems \cite{nambu2004,takesue2007}. We explain the key component of the attack: how Eve can control Bob's SNSPDs in the DPS-QKD system. Bob consists of an unbalanced Mach-Zehnder interferometer, and two detectors $\textrm{D}_0$ and $\textrm{D}_1$ (Fig.~\ref{fig:dps-bob-control}(a)). We assume that a properly implemented Bob will not accept clicks from both detectors for the duration of recovery after a click in one of the detectors, in order to avoid the detector deadtime and efficiency mismatch loopholes \cite{weier2011,makarov2006}. As illustrated above, the expected recovery is $\sim 40\,\nano\second$ long. Eve begins by applying to both detectors a laser pulse longer than the recovery time (Fig.~\ref{fig:dps-bob-control}(b)), with phase $\varphi$ changing in steps along the pulse such that its power splits equally to the two detectors. This pulse produces a double click at the beginning, which however can be timed to fall in between the bit slots and be discarded by Bob (the extra clicks may affect routines that adjust timing of Bob's acceptance windows, but note that attacks on such calibration routines are also possible \cite{jain2011}). Immediately after this long pulse, Eve applies a sequence of short pulses. Their phases are chosen to steer them primarily to one of the two detectors (similarly to \cite{makarov2008,lydersen2011}) and form hotspots in that detector only, keeping the comparator input voltage above the threshold. In the other detector, the voltage is allowed to fall below the comparator threshold. Then a pulse is applied and causes a click only in the detector that has recovered. Eve can end her control diagram here, or repeat the long pulse (as shown in Fig.~\ref{fig:dps-bob-control}(b)) and then make another controlled click. The total length of such chained control diagram producing several controlled clicks is limited by low-frequency cutoff of the RF components, and in the case of our setup can be up to $500\,\nano\second$. We remark that the short-pulsed parts of the diagram could in principle be replaced by a single phase-modulated long pulse, however short pulses may be easier to steer between Bob's detectors in case of sub-nanosecond $\Delta t$ used in the modern DPS-QKD systems \cite{takesue2007}.

Interferometers used for DPS-QKD are of a sufficiently good quality to allow Eve an extinction ratio of at least $20\,\deci\bel$ when routing her short pulses between the two Bob's detectors \cite{takesue2007}. Examination of the recovery traces in Fig.~\ref{fig:hotspot-during-recovery} and hotspot formation probabilities in Fig.~\ref{fig:deadtime-extension-detection-probabilities} suggests that the above control diagram will work. It should allow Eve to make clicks in Bob deterministically, or close to deterministically, in a wide range of comparator threshold voltages and $\Delta t$, even for $\Delta t = 100\,\pico\second$ \cite{takesue2007} or/and a threshold voltage above $20\,\milli\volt$. Eve should be able to vary the number of short pulses during the recovery to suit these system parameters, and still induce clicks in the correct detector most of the time.

While we did not have access to a complete DPS-QKD system to fully verify this Bob control method, we tested it experimentally by reproducing the expected power diagrams (optical power at $\textrm{D}_0$ and $\textrm{D}_1$ in Fig.~\ref{fig:dps-bob-control}(b)) at the single detector. We used $\Delta t = 5\,\nano\second$, and threshold setting of $11.6\,\milli\volt$ at the SR400 counter. We applied to the detector $2\,\milli\watt$ peak power, $53\,\nano\second$ long optical pulse, followed by $53\,\pico\second$ FWHM short optical pulses of varying energy. Measurement of the click probability while varying the short pulse energy showed that nearly perfect detector control ($< 0.005\%$ click probability in the wrong detector) would be achievable if Bob's interferometer in the DPS-QKD system had a reasonable $20\,\deci\bel$ extinction ratio, and good control ($< 1\%$ click probability in the wrong detector) would be possible at a very poor $10\,\deci\bel$ extinction ratio. The extinction ratio of Bob's interferometer determines how well Eve can suppress her short pulses from reaching the wrong detector, while making the target detector click with nearly unity probability. Jitter of the controllable click caused by the short pulse in the target detector was $250\,\pico\second$ FWHM, while that of the double click caused by the long pulse's leading edge was $170\,\pico\second$ FWHM.

One can notice that Eve would need to know Bob's detector parameters rather precisely to execute this attack. In modern cryptography, according to Kerckhoffs' principle \cite{kerckhoffs1883}, properties of equipment are assumed to be fully known to Eve. In practice, to learn the detector parameters, Eve might at first try to attack intermittently a few bits at a time (which would not raise the error rate noticeably) while varying her attack parameters, and watch the public discussion between Alice and Bob \cite{makarov2005}. One can also notice that Eve's intercept-and-resend equipment would introduce an insertion delay of at least some tens of $\nano\second$. However, photon's time-of-flight is not authenticated in today's implementations of QKD, and is not a part of the practical QKD protocols. Furthermore, in a fiber-optic line, Eve can easily cancel this insertion delay by shortcutting a part of the line between her intercept and resend units with a line-of-sight radio-frequency classical link \cite{makarov2005,gerhardt2011}.

\section{Discussion and conclusion}
\label{sec:conclusion}
The experimental results show that the control of this SNSPD is nearly perfect. Therefore, if this SNSPD were used in a QKD system, an eavesdropper could use bright illumination to capture the full raw and secret key, while introducing negligible errors. Installation of the eavesdropper is fully reversible: The detector survived the bright illumination with no signs of damage or deterioration \footnoterecall{fn:testing-restricted-to-one-sample}.

While the SNSPD is based on different physics than the APD single-photon detector, the similarity in how they can be controlled is startling. Latching the SNSPD using bright illumination can be considered as permanently blinding it, without the need for additional illumination to keep it blind. In the latched/blind state, the SNSPD exhibits the same superlinear response to bright trigger pulses as a blind APD. Likewise, controlling the SNSPD using deadtime extension is nearly identical to controlling the APD using deadtime extension: the only difference is that for this SNSPD the low-frequency cut-off of the RF components (and on a longer time scale the latching phenomenon) limits how long the deadtime can be extended.

Countermeasures against bright illumination attacks have been discussed extensively \cite{makarov2009,lydersen2010a,lydersen2010b,yuan2010,lydersen2010c,lydersen2011a,gerhardt2011,yuan2011,lydersen2011d}, and the conclusions are equally applicable to SNSPDs. The difference between public-key cryptography and QKD is that for the latter there exist security proofs. However, when the security is proved for systems with imperfections, models are used for the devices in the implementations. Even if this experiment is only performed on one device, \emph{the results show that the response deviates considerably from the models} in the simple security proofs that are usually employed \cite{lo1999,shor2000,gottesman2004}. There are more advanced security proofs that could allow such a response under certain conditions \cite{maroy2010,lydersen2011b}, but this would require discarding large amounts of the raw key to remove Eve's knowledge about the final key. For gated APD-based detectors, there is a proposal to bound the detector parameters by including a calibrated light source inside Bob, randomly testing, and thereby guaranteeing the single-photon sensitivity at random times \cite{lydersen2011a}. Another approach suggests to move detectors outside the secure devices and thus outside the security proof \cite{lo2011,braunstein2011}.

If one only wants to avoid these specific attacks, proposals for APD-based detectors \cite{makarov2009,lydersen2010a,lydersen2010b,yuan2010,lydersen2010c,lydersen2011a,gerhardt2011,yuan2011,lydersen2011d} should be equally efficient on SNSPDs, for instance an optical power meter at the entrance of Bob. In an installed QKD system, latching should be avoided either by an automated reset, or by including a shunt resistor in parallel with the nanowire \footnote{R.\ Hadfield, private communication (2011).}, but this does not guarantee that latching is precluded for all types of external input. Developing such specific countermeasures effective against \emph{specific,} known attacks is less than satisfactory, because this introduces an unproven extra assumption into the QKD security model that the countermeasure also eliminates all \emph{unknown} attacks exploiting the same loophole. Meanwhile, the difference between the device and its model in the QKD security proof remains. This approach would downgrade the level of QKD security model to that of public-key cryptography, which also includes unproven assumptions (of computational complexity).

As mentioned in the introduction, SNSPDs are still in their infancy, and therefore our findings might not apply to other detector designs. However, our findings clearly demonstrate that unless detector control is specially considered during design, SNSPDs may be controllable using bright illumination, just as their APD-based cousins. Furthermore, it could be possible to design the SNSPDs to be compliant with security proofs for QKD. The early stage of SNSPD technology is an excellent opportunity to avoid detector control vulnerability for future generations of SNSPDs. Designing hack-proof detectors will be crucial for the success of QKD.

\begin{acknowledgments}
We thank A.~Korneev, R.~Hadfield and V.~Burenkov for useful discussions. This work was supported by the Research Council of Norway (grant no.~180439/V30), Ontario Center of Excellence (OCE) and National Research Council of Canada (NSERC). L.L.\ and V.M.\ thank Institute for Quantum Computing for travel support during their visit to Waterloo.
\end{acknowledgments}

\end{document}